\documentclass{aa}  

\usepackage{natbib}
\bibpunct{(}{)}{;}{a}{}{,} 
\usepackage{mathptmx}
\usepackage{graphicx}
\usepackage[colorlinks=True,allcolors=blue]{hyperref}
\usepackage{txfonts}
\begin{document}

   \title{The ALPINE-ALMA [CII] Survey: the population of [CII]-undetected galaxies and their role in the $\mathrm{L_{[CII]}}$-SFR relation}
   \authorrunning{M. Romano et al.}
   \titlerunning{The role of non-detections in the $\mathrm{L_{[CII]}}$-SFR relation}

\author{M. Romano\thanks{E-mail: michael.romano@studenti.unipd.it}\inst{1,2,3}
\and
L. Morselli\inst{4}
\and
P. Cassata\inst{1,2}
\and
M. Ginolfi\inst{5}
\and
D. Schaerer\inst{6}
\and
M. B\'ethermin\inst{7}
\and
P. Capak\inst{8}
\and\\
A. Faisst\inst{9}
\and
O. Le Fèvre\thanks{Passed away.}\inst{7}
\and
J. D. Silverman\inst{10,11}
\and
L. Yan\inst{12}
\and
S. Bardelli\inst{13}
\and
M. Boquien\inst{14}
\and
M. Dessauges-Zavadsky\inst{6}
\and\\
S. Fujimoto\inst{15,16}
\and
N. P. Hathi\inst{17}
\and
G. C. Jones\inst{18,19}
\and
A. M. Koekemoer\inst{17}
\and
B. C. Lemaux\inst{20,21}
\and
H. M\'endez-Hern\'andez\inst{22}
\and\\
D. Narayanan\inst{23,24,25}
\and
M. Talia\inst{26,13}
\and
D. Vergani\inst{13}
\and
G. Zamorani\inst{13}
\and
E. Zucca\inst{13}
}

\institute{
Dipartimento di Fisica e Astronomia, Universit\`a di Padova, Vicolo dell'Osservatorio 3, I-35122, Padova, Italy
\and
INAF - Osservatorio Astronomico di Padova, Vicolo dell'Osservatorio 5, I-35122, Padova, Italy
\and
National Center for Nuclear Research, ul. Pasteura 7, 02-093 Warsaw, Poland
\and
Associazione Big Data, via Piero Gobetti 101, 40129 Bologna, Italy
\and
European Southern Observatory, Karl-Schwarzschild-Strasse 2, 85748, Garching, Germany
\and
Observatoire de Genève, Universit\'e de Genève, 51 Ch. des Maillettes, 1290 Versoix, Switzerland
\and
Aix Marseille Univ, CNRS, CNES, LAM, Marseille, France
\and
Infrared Processing and Analysis Center, California Institute of Technology, Pasadena, CA 91125, USA
\and
IPAC, California Institute of Technology, 1200 East California Boulevard, Pasadena, CA 91125, USA
\and
Kavli Institute for the Physics and Mathematics of the Universe, The University of Tokyo Kashiwa, Chiba 277-8583, Japan
\and
Department of Astronomy, School of Science, The University of Tokyo, 7-3-1 Hongo, Bunkyo, Tokyo 113-0033, Japan
\and
The Caltech Optical Observatories, California Institute of Technology, Pasadena, CA 91125, USA
\and
INAF - Osservatorio di Astrofisica e Scienza dello Spazio, via Gobetti 93/3, I-40129 Bologna, Italy
\and
Centro de Astronom\'ia (CITEVA), Universidad de Antofagasta, Avenida Angamos 601, Antofagasta, Chile
\and
Cosmic Dawn Center (DAWN), Jagtvej 128, DK22000 Copenhagen N, Denmark
\and
Niels Bohr Institute, University of Copenhagen, Lyngbyvej 2, DK2100 Copenhagen \O, Denmark
\and
Space Telescope Science Institute, 3700 San Martin Drive, Baltimore, MD 21218, USA
\and
Cavendish Laboratory, University of Cambridge, 19 J. J. Thomson Ave., Cambridge CB3 0HE, UK
\and
Kavli Institute for Cosmology, University of Cambridge, Madingley Road, Cambridge CB3 0HA, UK
\and
Department of Physics and Astronomy, University of California, Davis, One Shields Ave., Davis, CA 95616, USA
\and
Gemini Observatory, NSF’s NOIRLab, 670 N. A’ohoku Place, Hilo, Hawai’i, 96720, USA
\and
Insituto de F\'isica y Astronom\'ia, Universidad de Valpara\'iso, Avda. Gran Breta\~na 1111, 2340000 Valpara\'iso, Chile
\and
Department of Astronomy, University of Florida, 211 Bryant Space Sciences Center, Gainesville, FL 32611 USA
\and
University of Florida Informatics Institute, 432 Newell Drive, CISE Bldg E251, Gainesville, FL 32611
\and
Cosmic Dawn Center at the Niels Bohr Institute, University of Copenhagen and DTU-Space, Technical University of Denmark
\and
University of Bologna - Department of Physics and Astronomy “Augusto Righi” (DIFA), Via Gobetti 93/2, I-40129, Bologna, Italy
}

\date{Received May 2021}

 
\abstract
{The [CII] 158~$\mu$m emission line represents so far one of the most profitable tools for the investigation of the high-redshift galaxies in the early Universe. Being one of the brightest cooling lines in the rest-frame far-infrared regime of star-forming galaxies, it has been successfully exploited as a tracer of star-formation rate (SFR) in local sources. The picture is more complex at higher redshifts, where its usability in this context is still under investigation. Recent results from the ALMA Large Program to INvestigate [CII] at Early times (ALPINE) survey suggest that there is no (or weak) evolution of the L$\mathrm{_{[CII]}}$-SFR relation up to $z\sim6$ but their reliability is hampered by the presence of a large population of [CII] non-detected galaxies. In this work, we characterize the population of [CII] non-detections in ALPINE. By stacking their ALMA spectra, we obtain a signal detected at $\sim5.1\sigma$, resulting in a [CII] luminosity of $\mathrm{log(L_\mathrm{[CII]}}/\mathrm{L_{\odot}})$ $\sim7.8$. When combining this value with those from the [CII] detections, we find a $\mathrm{L_{[CII]}}$-SFR relation with a slope $b=1.14\pm0.11$, in agreement within the uncertainties both with the linear relation found in the local Universe, and with the previous findings from ALPINE at $z\sim5$. This suggests that the [CII] line can be considered a good tracer of star formation up to the distant Universe. Finally, we show that the galaxies of our sample that most deviate from the observed L$_\mathrm{[CII]}$-SFR relation could suffer from a less precise redshift estimation, perhaps artificially reducing their [CII] luminosity. In this respect, we claim that there is no evidence in favour of a deficit of [CII] content in high-z galaxies, in contrast with earlier studies.} 
  

\maketitle
%

\section{Introduction}
Over the last years, observations of the [CII] line emission at 158~$\mu$m rest-frame in galaxies have progressively improved to the point of being able to characterize the first sources of light during or near the epoch of cosmic Reionization (e.g., \citealt{Wagg12,Carilli13,Wang13,Capak15,Pentericci16,Carniani18,Smit18,Hashimoto19,Matthee19,Bakx20,Harikane20,LeFevre20,Bouwens21,Fudamoto21}).

The widespread interest in detecting the [CII] emission from local and distant galaxies is highly justified. This is the brightest line arising from the rest-frame far-infrared (FIR) spectra of star-forming galaxies (SFGs), representing one of the main coolants of their interstellar medium (ISM; e.g., \citealt{Stacey91,Carilli13}). The bulk of its emission originates from photo-dissociation regions (PDRs; \citealt{Hollenbach99}), possibly tracing the formation of new stars from giant molecular clouds. However, given its low ionization potential (11.3 eV, compared to the 13.6 eV of neutral hydrogen), this line deserves a thoughtful physical interpretation as it can also trace the diffuse neutral medium (e.g., \citealt{Wolfire03,Vallini15}), the molecular (e.g., \citealt{Zanella18,Dessauges20}) and ionized (e.g., \citealt{Cormier12}) gas. Furthermore, there is evidence that [CII] can reasonably trace the total HI content of galaxies \citep{Heintz21}. The continuum surrounding this line lies close to the peak of the FIR dust-emission, making detection easier and helping to constrain the total FIR luminosity and obscured star formation (e.g., \citealt{Gruppioni20,Khusanova20,Pozzi21}). Moreover, the [CII] emission can provide important information on a variety of ISM properties, such as the star-formation rate (SFR; e.g., \citealt{DeLooze14,Olsen17}), the presence of outflows (e.g., \citealt{Gallerani18,Ginolfi20}), and the kinematics of the ISM (e.g., \citealt{Jones21,Romano21}). Therefore, the comparison of [CII] observations with simulations is mandatory to disentangle the diverse processes that take place in galaxies, in order to understand how they shape the observed morphology and kinematics of the line (e.g., \citealt{Vallini17,Ferrara19,Kohandel19}).

In this context, the ALMA Large Program to INvestigate [CII] at Early times (ALPINE) survey \citep{Bethermin20,Faisst20,LeFevre20} has recently provided the first statistically significant sample of high-redshift \textit{normal} galaxies (i.e., lying along a well-defined \lq\lq main-sequence\rq\rq{} of galaxies, with a relatively tight dispersion ($<0.3$~dex) in the SFR versus stellar mass plane; e.g., \citealt{Noeske07,Rodighiero11,Speagle14}) detected in [CII] at the end of the reionization epoch ($4.4<z<5.9$). ALPINE observed a sample of 118 SFGs selected in UV and with redshifts spectroscopically confirmed by previous campaigns \citep{LeFevre15,Tasca17,Hasinger18} in order to ensure precise detections of the [CII] line. In particular, considering as detections those galaxies with a [CII] emission $\geq3.5\sigma$ (corresponding to a 95\% sample purity), ALPINE reached a successful rate of 64\%, resulting in 75 detections and 43 non-detections \citep{Bethermin20,LeFevre20}.

\cite{Bethermin20} first investigated the $\mathrm{L_{[CII]}}$-SFR relation by taking advantage of the ALPINE continuum-detected galaxies. They computed the average SFRs in different [CII] luminosity bins as the sum of the UV rest-frame data \citep{Faisst20} and of the mean obscured SFRs derived through the stacking of the continuum data (i.e., $\rm SFR_{total}=SFR_{UV}+SFR_{IR}$). Their results are in good agreement with the local and predicted relations. Then, \cite{Schaerer20} took advantage of the full ALPINE sample (including both [CII]-detected galaxies and upper limits on non-detections; see Section \ref{sec:results}) to study the evolution of the $\mathrm{L_{[CII]}}$-SFR relation over cosmic time, and to understand if the [CII] line is a good tracer of SFR at high redshift as it is in the local Universe (e.g., \citealt{DeLooze14,Pineda14}). They found that the [CII] luminosity of the ALPINE galaxies scales linearly with their total SFRs (as traced by the sum of UV and IR contributions; see \citealt{Schaerer20}), with a slight steepening of the slope depending on the [CII] non-detections upper limits used \citep{Bethermin20}. However, to fully establish the connection between $\mathrm{L_{[CII]}}$ and SFR in distant galaxies a more in depth investigation of the ALPINE non-detections is needed.  

In this work, we derive the average properties of the population of [CII] undetected galaxies in ALPINE through line stacking and use them to investigate the [CII] as a tracer of SFR at high redshift, and to put additional constraints on the already thoroughly studied $\mathrm{L_{[CII]}}$-SFR relation \citep{Schaerer20}. 

The structure of the paper is the following: in Section \ref{sec:data} we introduce the available data and observations used to compute the [CII] line profile resulting from the average population of non-detections. The procedure adopted to obtain the stacked line is described in Section \ref{sec:stacking}. The results are reported in Section \ref{sec:results} and discussed in Section \ref{sec:discussion}, respectively. Summary and conclusions are provided in Section \ref{sec:summary}. 

Throughout this work, we adopt a $\mathrm{\Lambda-CDM}$ cosmology with $H_0=70$~km~s$^{-1}$~Mpc$^{-1}$, $\mathrm{\Omega_m}=0.3$ and $\mathrm{\Omega_\Lambda}=0.7$. Furthermore, we use a \cite{Chabrier03} initial mass function (IMF).

\begin{figure}
    \begin{center}
	\includegraphics[width=\columnwidth]{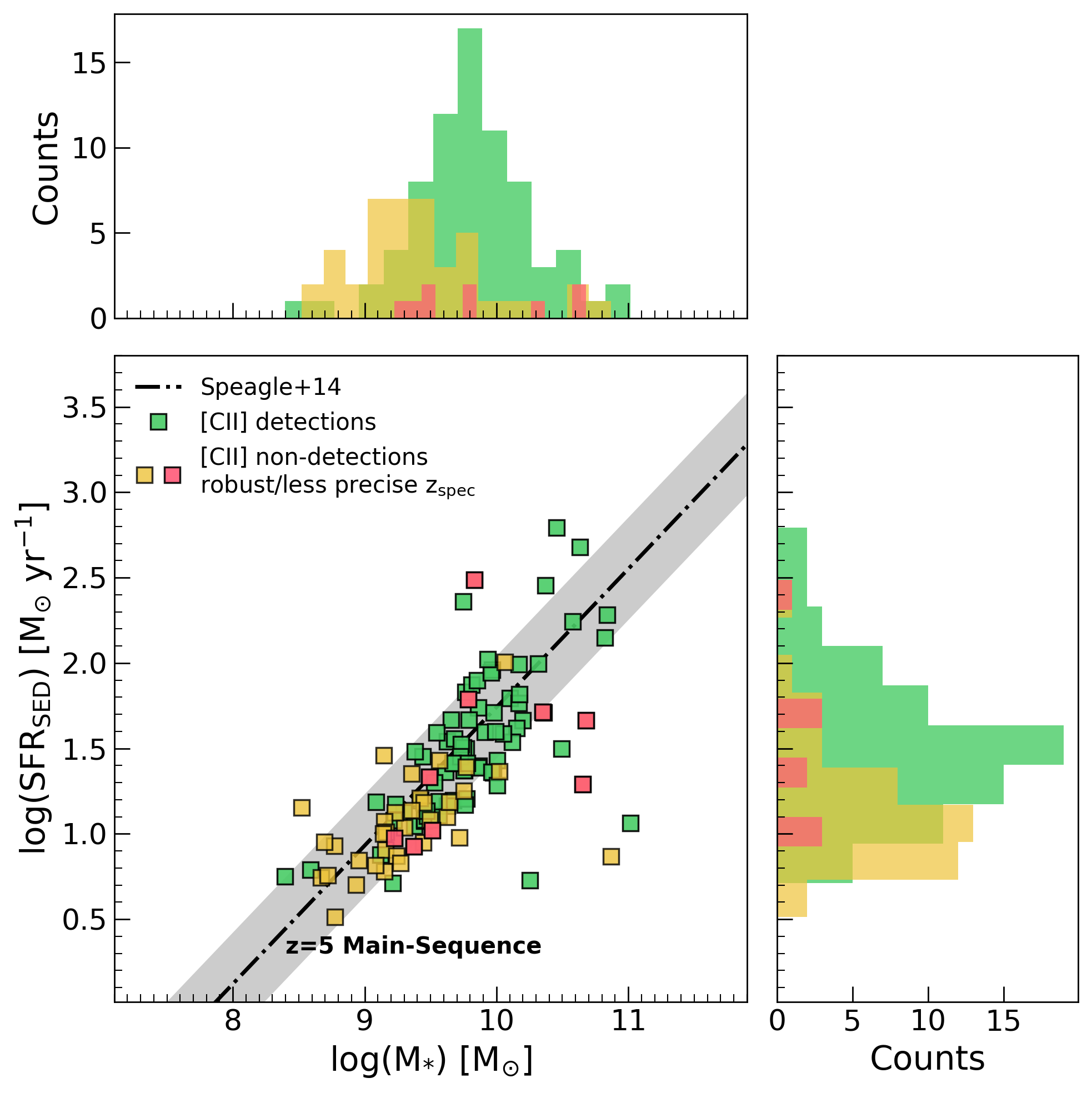}
	\end{center}
    \caption{SFRs vs stellar masses of the ALPINE [CII]-detected (green squares) and undetected (yellow squares) galaxies. Red symbols represent [CII] non-detections with less precise spectroscopic redshift (see Section \ref{sec:UV_data}). The main-sequence of star-forming galaxies by \cite{Speagle14} is shown as the black dot-dashed line, with its $\pm0.3$~dex width represented by the grey shaded region. The top and right panels report the distributions in stellar mass and SFR, respectively, for both the detections and non-detections.}
    \label{fig:MS}
\end{figure}

\begin{figure*}[t!]
    \begin{center}
	\includegraphics[width=1.7\columnwidth]{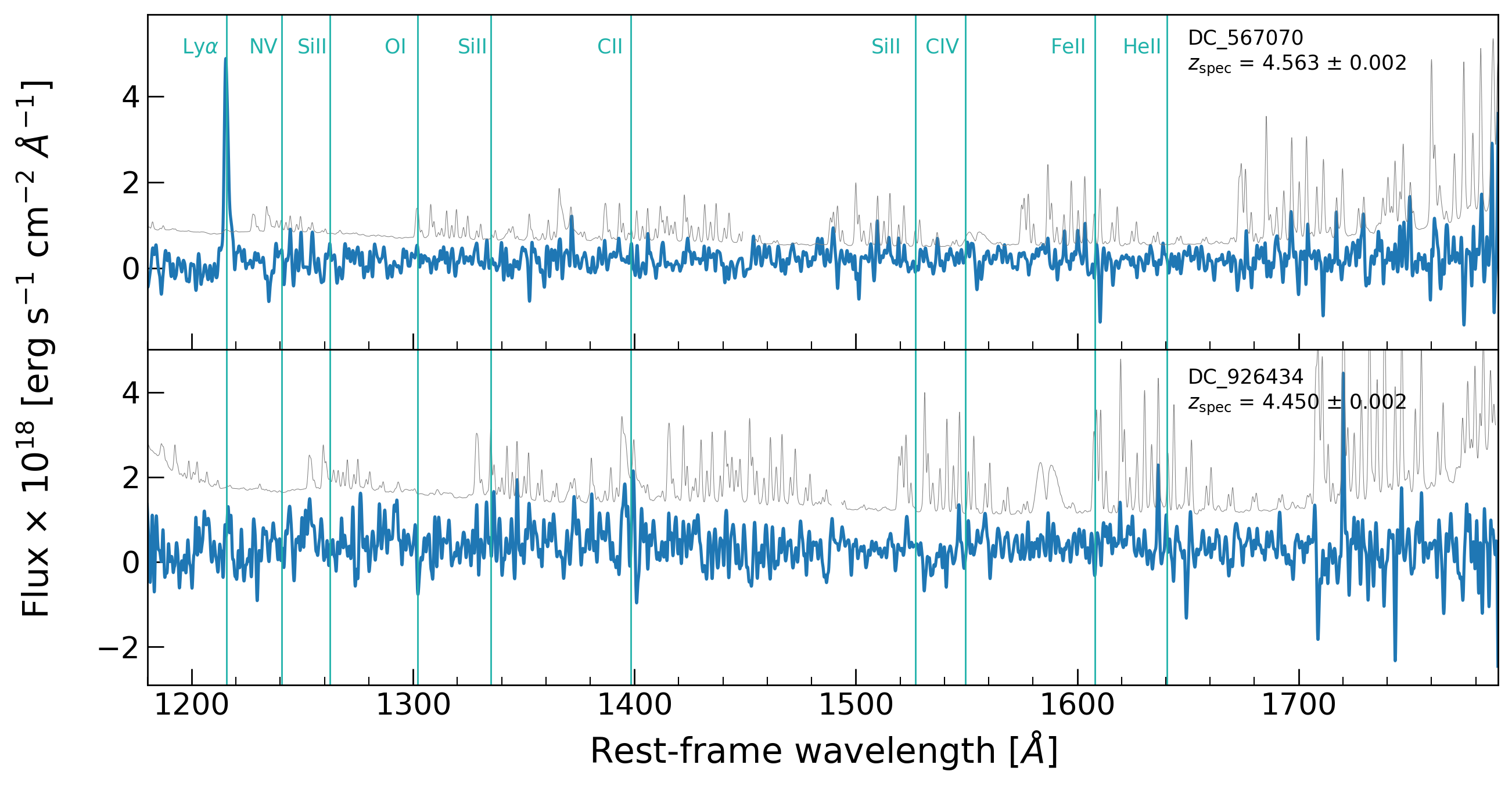}
	\end{center}
    \caption{Examples of optical spectra of two [CII] non-detections with robust (top) and less precise (bottom) rest-frame UV spectroscopic redshift. The typical redshift uncertainty due to the spectral resolution (i.e., $\mathrm{R\sim2700}$ from \citealt{Hasinger18}) of the observations is also shown. Both panels report UV emission and absorption features, such as the Ly$\alpha$ line or the ISM Si\,{\sc II}, C\,{\sc IV} and He\,{\sc II} absorption lines. The spectra (in blue) and the noise (in gray) are smoothed with a Gaussian filter with size of $2~\AA$ for a better visualization of the emission/absorption features.}
    \label{fig:spectra_ex}
\end{figure*}

\section{Data and observations}\label{sec:data}
\subsection{Multi-wavelength and ALMA data}\label{sec:multiwave}
The ALPINE survey was designed to observe the [CII] line at 158~$\mu$m rest-frame and the surrounding FIR continuum emission from a sample of 118 SFGs at $4.4<z<5.9$, avoiding the redshift range $4.6<z<5.1$ due to a low-transmission atmospheric window. The campaign spent $\sim70$~hours of observation in ALMA Band 7 (275-373~GHz) during cycles 5 and 6. The targets are drawn from the well-studied Cosmic Evolution Survey (COSMOS; \citealt{Scoville17a,Scoville17b}) and Extended Chandra Deep Field South (E-CDFS; \citealt{Giavalisco04,Cardamone10}) fields and have been observed in large optical/NIR spectroscopic campaigns such as VIMOS UltraDeep Survey (VUDS; \citealt{LeFevre15,Tasca17}) and DEIMOS 10K Spectroscopic Survey \citep{Hasinger18}. They are selected in the rest-frame UV ($\mathrm{L_{UV}>0.6~L^{*}}$) and lie on the main-sequence, thus being representative of the average population of star-forming galaxies at $z\sim5$ (e.g., \citealt{Speagle14}). A wealth of multi-wavelength data is available for these sources, from the UV to the near-IR (e.g., \citealt{Koekemoer07,Sanders07,McCracken12,Guo13,Nayyeri17}), allowing us to recover physical quantities such as stellar masses ($\mathrm{9\lesssim log(M_{*}/M_{\odot})\lesssim11}$) and star-formation rates ($\mathrm{1\lesssim log(SFR/M_{\odot}yr^{-1})\lesssim3}$) through spectral energy distribution (SED)-fitting \citep{Faisst20}. Data are also available in the X-ray and radio bands (e.g., \citealt{Hasinger07,Smolcic17}).

The ALMA data-cubes were reduced and calibrated using the standard Common Astronomy Software Applications (CASA; \citealt{McMullin07}) pipeline. Each cube was continuum-subtracted in the $uv$-plane in order to obtain line-only cubes with channel width of $\sim25~\mathrm{km~s^{-1}}$ and beam size of $\sim1''$ (with a pixel scale of $\sim0.15''$; \citealt{Bethermin20}). A line search algorithm was then applied to each continuum-subtracted cube resulting in 75 [CII] detections (signal-to-noise ratio; $\mathrm{S/N>3.5}$) and 43 non-detections. In Figure \ref{fig:MS} we show the distributions of [CII]-detected and undetected galaxies along the $z\sim5$ main-sequence of SFGs. The stellar masses and SFRs are those from the SED-fitting \citep{Faisst20}. As evident, the ALPINE non-detections lie on the bottom-left side of the main-sequence, at lower stellar masses and SFRs with respect to those detected in [CII] (except for a few massive sources below the sequence).

For a more in-depth description of the overall ALPINE survey, the observations, data processing, and multi-wavelength analysis see \cite{LeFevre20}, \cite{Bethermin20} and \cite{Faisst20}, respectively.

\subsection{Rest-frame UV spectroscopic data}\label{sec:UV_data}
The 118 ALPINE galaxies have confirmed rest-frame UV spectroscopic redshifts from the VUDS \citep{LeFevre15,Tasca15} and DEIMOS 10K \citep{Hasinger18} surveys. These are obtained both from the Ly$\alpha$ line and from UV rest-frame ISM absorption lines. However, these features are not always the best tracers of the systemic redshift of a source. The Ly$\alpha$ emission line is typically redshifted (with respect to the systemic velocity) because of the resonant scattering of the Ly$\alpha$ photons (e.g., \citealt{Dijkstra14,Verhamme15}). On the opposite, ISM lines are usually blueshifted, suggesting the presence of outflowing gas (e.g., \citealt{Steidel10,Steidel18}). The [CII] line is not affected by this kind of issues and, in principle, it can be used to stack together the ALMA spectra of the ALPINE non-detections to search for a significant signal. Moreover, it is not absorbed by dust and can be observed across the entire galaxy, resulting in a better tracer of the systemic redshift than optical nebular lines (e.g., \citealt{Cassata20,Faisst20}).

\begin{figure*}
    \begin{center}
	\includegraphics[width=\textwidth]{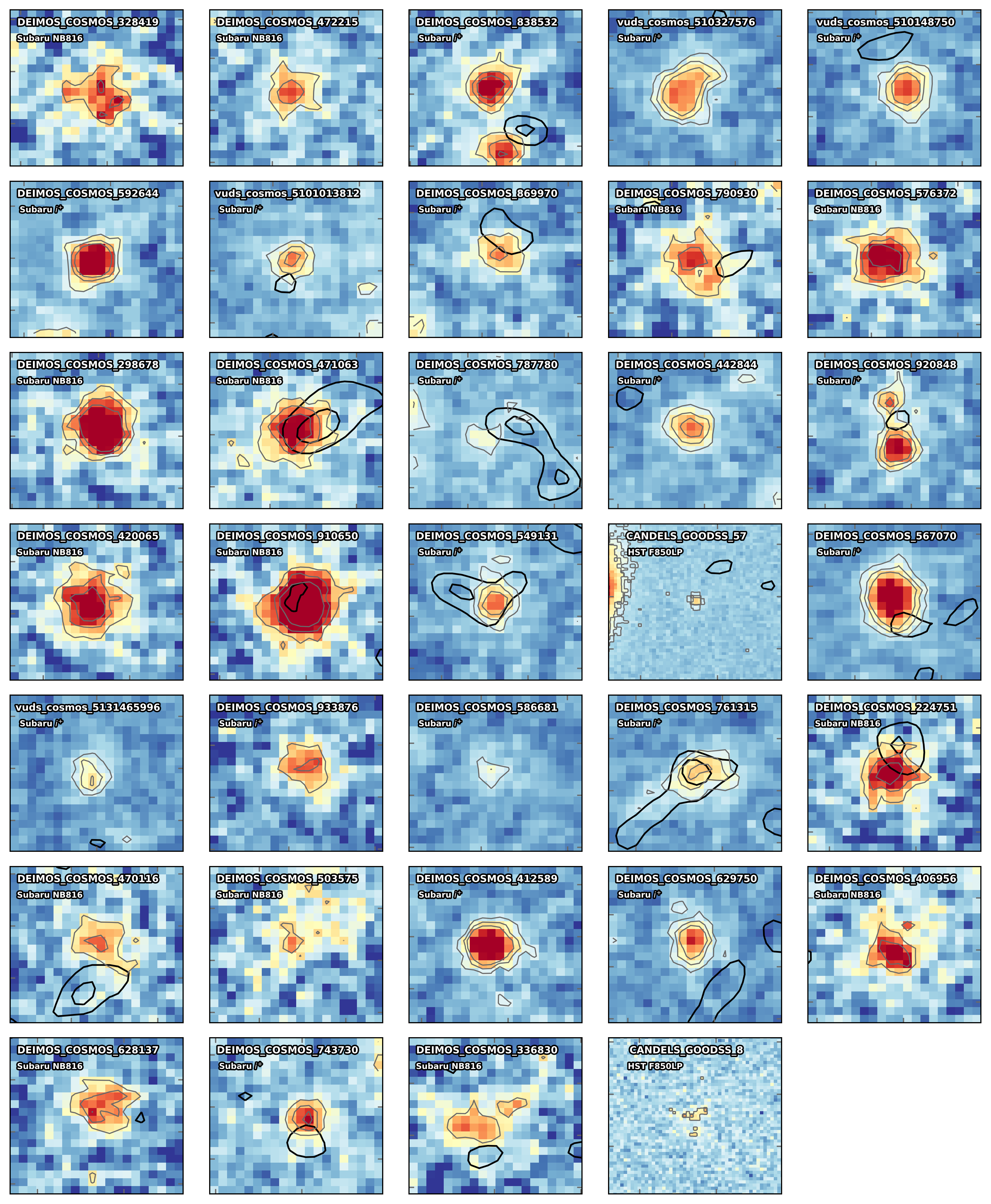}
	\end{center}
    \caption{Cutouts of the 34 [CII] non-detections centered on their optical positions. Depending on the method used to pre-select each galaxy before the spectroscopic observation, i.e., Lyman-break, Ly$\alpha$, or $i$-band dropout selection, we show the emission from the Subaru broad-band $i^{+}$ and narrow-band NB816 filters \citep{Taniguchi07,Taniguchi15,Laigle16}, or from the HST/ACS F850LP \citep{Giavalisco04}, respectively (see \citealt{Faisst20}). Each image is $3''\times3''$ wide. Grey contours show 3, 5, and 7$\sigma$ optical emission. Black contours (if present) represent 2 and 3$\sigma$ ALMA [CII] emission. The name of each non-detection, as well as the filter used, are reported in the upper-left corner of the corresponding cutout.}
    \label{fig:cutout1}
\end{figure*}

We do not have the systemic redshift information for the 43 ALPINE non-detections, therefore we can just rely on their UV redshifts. In particular, we need to know the UV spectroscopic redshifts of our sources with good accuracy, in order to exclude objects with less precise redshift estimates for which the [CII] line could lie outside of the ALMA observational window or that could alter the $\Delta v_{\mathrm{Ly}\alpha}$ statistics in our stacking analysis (see Section \ref{sec:stacking}). For this reason, we visually inspected the optical spectra of the non-detections. We found that 34 out of 43 sources present multiple high S/N spectral features, all of them showing a prominent Ly$\alpha$ line in emission, allowing for a precise and accurate estimate of their spectroscopic redshifts. The remaining 9 galaxies have very weak or no Ly$\alpha$ in emission\footnote{We note that there are many factors that could absorb and/or decrease the Ly$\alpha$ emission, like the presence of dust and its geometrical distribution, or the clumpiness of the ISM (e.g., \citealt{Dijkstra14,Messa19,Cassata20}).} and less prominent and sharp UV absorption lines. Although it is likely that the redshift is generally accurate for these sources as well, we decide to exclude them from our stacking analysis, in order not to include additional uncertainties due to possible strong [CII] offsets with respect to the expected position of the line. Figure \ref{fig:spectra_ex} reports an example of UV/optical spectra of two [CII] non-detections with a robust and less precise spectroscopic redshift, respectively. In the first case, the galaxy shows a clear Ly$\alpha$ line in emission and some other possible absorption features at longer wavelengths that provide a precise estimate of the spectroscopic redshift. On the other hand, the spectrum of the second source is quite noisy, with fewer recognizable spectral features. We thus obtain a final sample of 34 [CII] non-detections. The rest-frame UV and ALMA emissions of these galaxies are shown in Figure \ref{fig:cutout1}, where it is evident that no significant [CII] signal is present in these sources\footnote{A few maps show a 3$\sigma$ [CII] emission in spatial coincidence with the rest-frame UV emission. Although these sources present signs of significant [CII] content, their were classified as non-detections given their S/N smaller than 3.5$\sigma$ (see Section \ref{sec:multiwave}).}. The 9 galaxies with less precise $z_\mathrm{spec}$ are also reported for completeness in Figure \ref{fig:MS} as red squares. 

For these reasons, in the following analysis, we make use of the redshift obtained from the peak of the Ly$\alpha$ line, and then correct it for the observed velocity offsets between Ly$\alpha$ and the systemic velocity traced by [CII] ($0<\Delta v_{\mathrm{Ly}\alpha}<400$~km~s$^{-1}$) for the ALPINE detections \citep{Cassata20}.

\section{Stacking of non-detections}\label{sec:stacking}
We proceed with a mean stacking of the ALMA spectra of the 34 ALPINE [CII] non-detections to search for a signal emerging from the noise of individual galaxies. 

In principle, the stacking analysis can be done both in the image and uv planes. In the first case, it is possible to perform both 1D and 3D stacking through the extraction of individual spectra and by exploring the full data-cubes, respectively (e.g., \citealt{Bischetti19,Stanley19}). 2D stacking can be performed as well, by aligning and summing intensity maps of different sources (e.g., \citealt{Mendez20}). The stacking in the uv plane is instead based on the analysis of the visibility data associated to the emission line in the Fourier space, before the imaging process (e.g., \citealt{Fujimoto19,Carvajal20}). \cite{Lindroos15} made use of several simulated uv continuum data sets to compare the performances of image and uv stacking. They found that the two procedures yield similar results within the uncertainties, although in some cases stacking in the uv plane could lead to higher S/N, also avoiding possible issues during the deconvolution process from the Fourier to the image domain. \cite{Mendez20} also made a comparison between the image and uv stacking of CO emission lines in a sample of 27 low-$z$ SFGs drawn from the Valpara\'iso ALMA/APEX Line Emission Survey (VALES; \citealt{Villanueva17,Cheng18}). They retrieved similar flux densities with both methods, with the uv stacking producing similar or slightly higher S/N as compared to the image one, depending on the brightness of the line. Regardless of the adopted method, all the above techniques need a good knowledge of the systemic redshift of the source. Furthermore, to select the spectral channels including the emission line in the case of uv stacking or for the production of the intensity maps, an estimate of the full width at half maximum (FWHM) of each source is also required. In the analysis of the ALPINE [CII] undetected galaxies, we do not know at which frequency the expected line is (because of the observed velocity offset between Ly$\alpha$ and [CII]), and we do not have any information on the FWHM of each source. For these reasons, and considering that both image and uv planes provide similar results, we decide to proceed with a 1D spectral stacking of the [CII] non-detections starting from the image plane.\\

\begin{figure}[t]
    \begin{center}
	\includegraphics[width=\columnwidth]{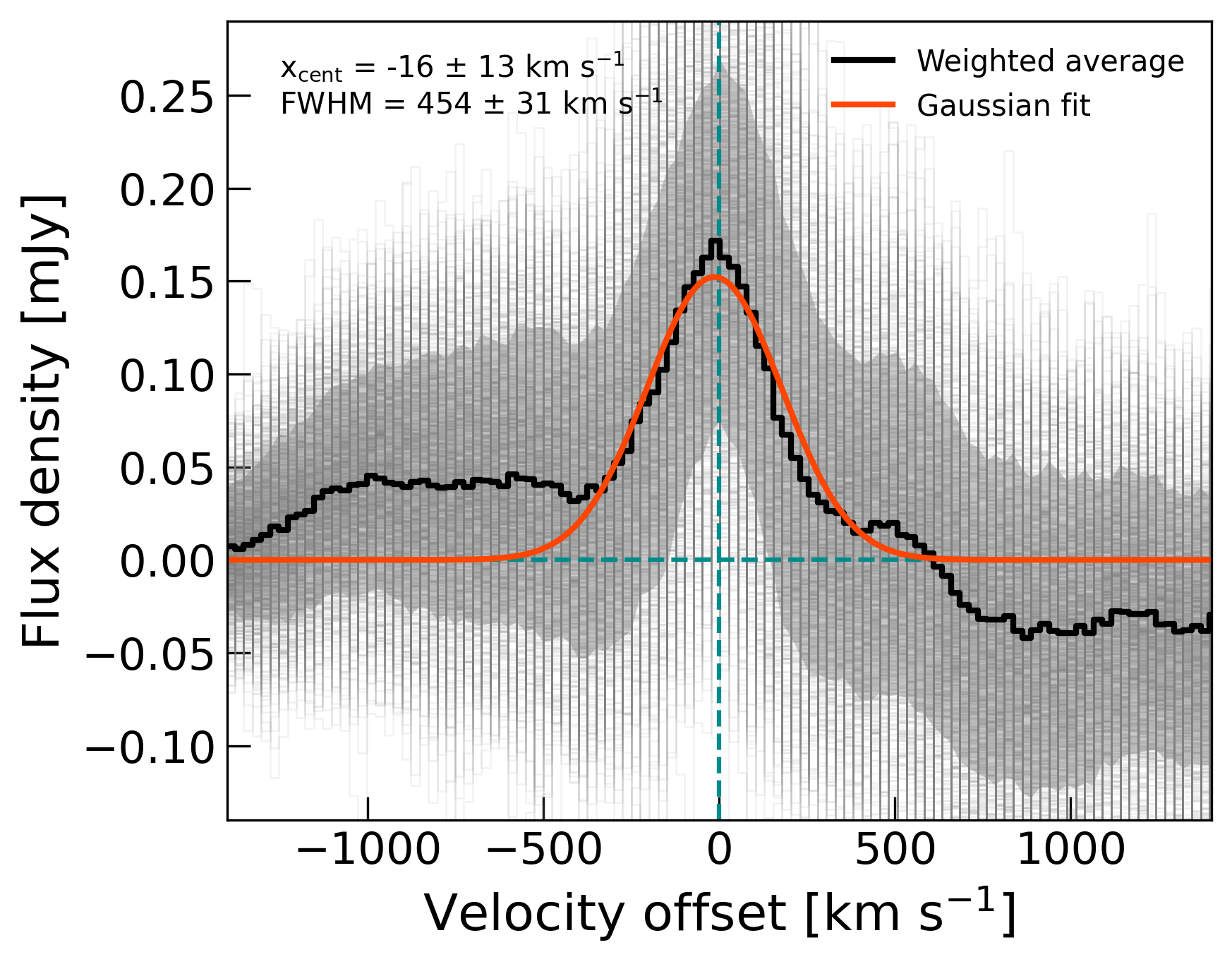}
	\end{center}
    \caption{Average spectrum of the ALPINE [CII] non-detections weighted for the S/N of the individual stacked spectra with $\mathrm{FWHM}\leq400$~km~s$^{-1}$ (solid black line). The thin lines represent the individual realizations after taking into account the observed shift between the [CII] and Ly$\alpha$ lines \citep{Cassata20}. The solid red line represents the Gaussian fit on the average line profile. The shaded area shows the uncertainty associated to the average line profile as taken from the 16th and 84th percentiles of the 1000 spectra distribution. The dashed dark cyan lines mark the zero flux and velocity offset levels. The centroid and FWHM computed from the Gaussian fit on the average line profile are shown on the top left corner.}
    \label{fig:stack}
\end{figure}

In particular, we extract each spectrum from the original data-cubes of the ALPINE data release 1 (DR1; \citealt{Bethermin20})\footnote{DR1 data are available at \url{https://cesam.lam.fr/a2c2s/data_release.php}.}, within a fixed aperture of 1'' of radius (defining the central regions of the ALPINE targets; see \citealt{Bethermin20}) centered at the optical position of the source\footnote{The coordinates of each galaxy are taken either from the COSMOS2015 \citep{Laigle16} or 3D-HST \citep{Brammer12,Skelton14} catalogs, depending on which field (COSMOS or E-CDFS) the source is located (see \citealt{Faisst20} for more information).}, and covering a spectral range of $\sim3$~GHz around the expected peak frequency of the emission line (as traced by the Ly$\alpha$ emission). At first, we use the Ly$\alpha$-based spectroscopic redshifts of the non-detections to align them to the same reference frame, and then we stack them together. Because the Ly$\alpha$ line is typically shifted to the red relative to the systemic redshift as defined by the [CII] line \citep{Cassata20}, it is likely that our stacked emission is offset from the systemic by some amount. Further, the velocity offset between Ly$\alpha$ and systemic is not constant, but is rather a complicated function of various physical conditions within a given galaxy (e.g., \citealt{Erb04,Pentericci16,Marchi19}), meaning that the signal recovered in a stack where these offsets were not accounted for would be broadened and dampened. For these reasons, we apply to each source a spectral offset randomly extracted from the $\Delta v_{\mathrm{Ly}\alpha}$ distribution obtained by \cite{Cassata20} from a sub-sample of the ALPINE detections, and then compute the mean stack on the shifted spectra. Certainly, by drawing random values out of the $\Delta v_{\mathrm{Ly}\alpha}$ we expect to introduce an artificial broadening of the stacked line, as well. In particular, the further (nearer) the rest-frame UV spectroscopic redshift (as traced by the Ly$\alpha$ line) from the systemic one (as traced by the [CII] line), the wider (narrower) the stacked line. To avoid this issue, we compute each time a Gaussian fit on the stacked line, estimating its full width at half maximum as $\mathrm{FWHM}=2.355\sigma$, where $\sigma$ is the standard deviation of the Gaussian. For each stacked spectrum, we then require that $\mathrm{FWHM}\leq400$~km~s$^{-1}$, which defines the 84th percentile of the observed FWHM distribution of the ALPINE [CII] detections (see \citealt{Bethermin20}), until reaching 1000 realizations. 

We show the result of this procedure in Figure \ref{fig:stack}, where the grey histograms represent the spectra that satisfy the above requirements on the FWHM. To produce the averaged line profile, we weight each realization of the stacked spectra in different ways. We assume $w=1$ in case of no weighting, $w=1/\sigma_{rms}^2$ for an inverse-variance weighting (where $\sigma_{rms}$ is the standard deviation of each stacked spectrum)\footnote{We check that the integrated flux and S/N of the final average stacked line do not change significantly if we weight the individual spectra by their observed rms in each realization. In the latter case, we obtain indeed a slightly lower flux (resulting in a $\sim0.1$~dex lower [CII] luminosity) and similar S/N if compared to those computed with our method, i.e., by directly weighting the stacked spectra of each realization for their rms. The two methods provide comparable results likely because of the distributions of the individual rms of our 34 spectra, that are quite constant in both of the two ALPINE redshift bins of our sample, providing similar weights to each source in the stack.}, and $w=\mathrm{S/N}$ to weight each stacked spectrum by its S/N. In the latter case, we estimate the S/N as the ratio between the peak flux of the stacked line and the standard deviation of the spectral channels at velocities greater and smaller than $\pm 600$~km~s$^{-1}$ from the peak, in order to avoid contamination from the stacked emission line. The line profiles obtained by applying these different weightings are similar to each other, although resulting in a slightly higher S/N for the case with $w=\mathrm{S/N}$. For this reason, we decide to show in Figure \ref{fig:stack} only the average stacked spectrum obtained by weighting each realization by its S/N. 

We then check the effect of the random extraction on the final average stacked profile, by making the following test. We select a sub-sample of [CII] detections having, in addition to the [CII]-traced systemic redshift, also the redshift measurements from Ly$\alpha$. We first stack together the ALMA spectra basing on their systemic redshifts (i.e., without applying any spectral offset), to obtain the real FWHM and flux of the line. Then, we repeat the stacking by using our method, i.e., by extracting random shifts from the $\Delta v_\mathrm{Ly\alpha}$ distribution, and compare the two results. We find that the random extraction method produces a $\sim2$ times broader FWHM than the true value computed by stacking the spectra at their correct wavelengths. At the same time, there is no significant difference in the retrieved S/N and integrated fluxes. As an additional check, we re-do the entire stacking procedure by considering, for each source, the mean value of the Ly$\alpha$-[CII] offset distribution found by \cite{Cassata20} (i.e., $\sim183~\mathrm{km~s^{-1}}$), obtaining a FWHM $\sim150~\mathrm{km~s^{-1}}$ narrower than what previously found. On the other hand, the integrated flux and S/N of the resulting line are slightly lower, but still comparable, than those computed by assuming the random $\Delta v_{\mathrm{Ly}\alpha}$ offsets. This further suggests that the random extraction has only a significant effect on the FWHM of the stacked line. Therefore, we are confident that the stacking method used in this work does not under/overestimate the true value of the [CII] luminosity of the non-detections population under study.

By fitting the stacked line profile with a Gaussian function, we find the signal peaking at $\mathrm{x_{cent}}=-16\pm13$~km~s$^{-1}$, consistent with the systemic velocity traced by the [CII] line, and having $\mathrm{FWHM}=454\pm31$~km~s$^{-1}$. The computed S/N is $\sim5.1$, revealing the presence of an underlying population of [CII] emitters likely suffering from low S/N (in terms of [CII] emission) possibly caused by the low SFRs and stellar masses which characterize them (see Figure \ref{fig:MS}). Following \cite{Solomon92}, we also compute $\mathrm{L_{[CII]}}$ as 
\begin{equation}
    \mathrm{L_{[CII]}} = 1.04 \times 10^{-3}~S_\mathrm{[CII]}~\Delta v~D_L^2~\nu_{obs}~[\mathrm{L_\odot}],
    \label{eq:line_lum}
\end{equation}
where $S_\mathrm{[CII]}~\Delta v$ is the velocity-integrated line flux in units of Jy~km~s$^{-1}$, $D_L$ is the luminosity distance in Mpc at the median redshift of the sample (i.e., $z_\mathrm{med}=5.52$), and $\nu_{obs}$ is the observed peak frequency in GHz.

To attribute an uncertainty on the velocity-integrated flux (and hence on the [CII] luminosity) we perform a delete-$d$ jackknife resampling \citep{Shao89}. We produce 500 jackknife realizations following the same method described above, but randomly removing each time 20\%\footnote{This percentage represents a good compromise in using the jackknife method to estimate the error on our measurements while still having a reasonable number of sources to stack at each time \citep{Efron82}.} of the 34 spectra in the parent sample and computing again the integrated flux of the stacked spectra. The error on $S_\mathrm{[CII]}~\Delta v$ is then computed from the 16th and 84th percentiles of the fluxes distribution of the 500 jackknife realizations. As a result, we obtain $S_\mathrm{[CII]}~\Delta v = (73 \pm 11)\times 10^{-3}$~Jy~km~s$^{-1}$.

We thus find $\mathrm{log(L_\mathrm{[CII]}}/\mathrm{L_{\odot}}) = 7.8\pm0.3$, where the error is computed by propagating the uncertainty on the [CII] flux and the redshift on Equation \ref{eq:line_lum}. We test that this result is not significantly different from the one obtained by removing the constraint on the FWHM of the stacked spectra, which thus affects only the width of the line.

Furthermore, given the wide range spanned by the [CII] non-detections both in stellar mass (i.e., $\mathrm{9\lesssim log(M_{*}/M_{\odot})\lesssim11}$) and star-formation rates (i.e., $\mathrm{0.5\lesssim log(SFR/M_{\odot}yr^{-1})\lesssim2.0}$; see Figure \ref{fig:MS}), we decide to split our sample in two bins of SFR to check if the above result from stacking is dominated by the most massive and star-forming sources. We use the median SFR of the sample ($\mathrm{log(SFR_{TOT,med}/M_{\odot}~yr^{-1}})=1.16$; as defined in Section \ref{sec:results}) to make two bins composed of 17 objects each, repeating for both of them the previously described stacking procedure. We do not find any significant emission in the low-SFR bin. On the contrary, a S/N $\sim6.8$ emission is produced in the high-SFR bin, suggesting that most of the signal in Figure \ref{fig:stack} is due to the most massive SFGs. By adopting the above-described stacking procedure on the whole sample, we find $\mathrm{log(L_\mathrm{[CII]}}/\mathrm{L_{\odot}}) = 8.0\pm0.2$ in the high-SFR bin. In the lower bin with no detection, we consider instead the average of the [CII] non-detections with robust spectroscopic redshift, computed as the addition in quadrature of the corresponding [CII] upper limits divided by the square root of the number of sources in the bin (e.g., \citealt{Cohen88}), finding $\mathrm{log(L_\mathrm{[CII]}}/\mathrm{L_{\odot}}) \lesssim 7.8$. We show these results in Figure \ref{fig:[CII]sfr}, along with the previous result from the stacking of all the [CII] non-detections.

Finally, Figure \ref{fig:stack} shows that the average [CII] profile presents a negative (positive) continuum at the right (left) of the line. By analyzing each spectrum individually, we attribute this behaviour to different sources that present a negative (positive) continuum at positive (negative) velocity offset with respect to the expected line. However, we believe that this issue is only due to statistical fluctuations originated from the small size of the sample, rather than to calibration problems in the ALMA data reductions or physical processes in these sources. This is also confirmed by the dispersion of the 1000 realizations of the stacked line in the figure, which shows that the above-mentioned continuum levels are not significant and consistent with zero. 

\section{Results}\label{sec:results}
\begin{figure}[t!]
    \begin{center}
	\includegraphics[width=\columnwidth]{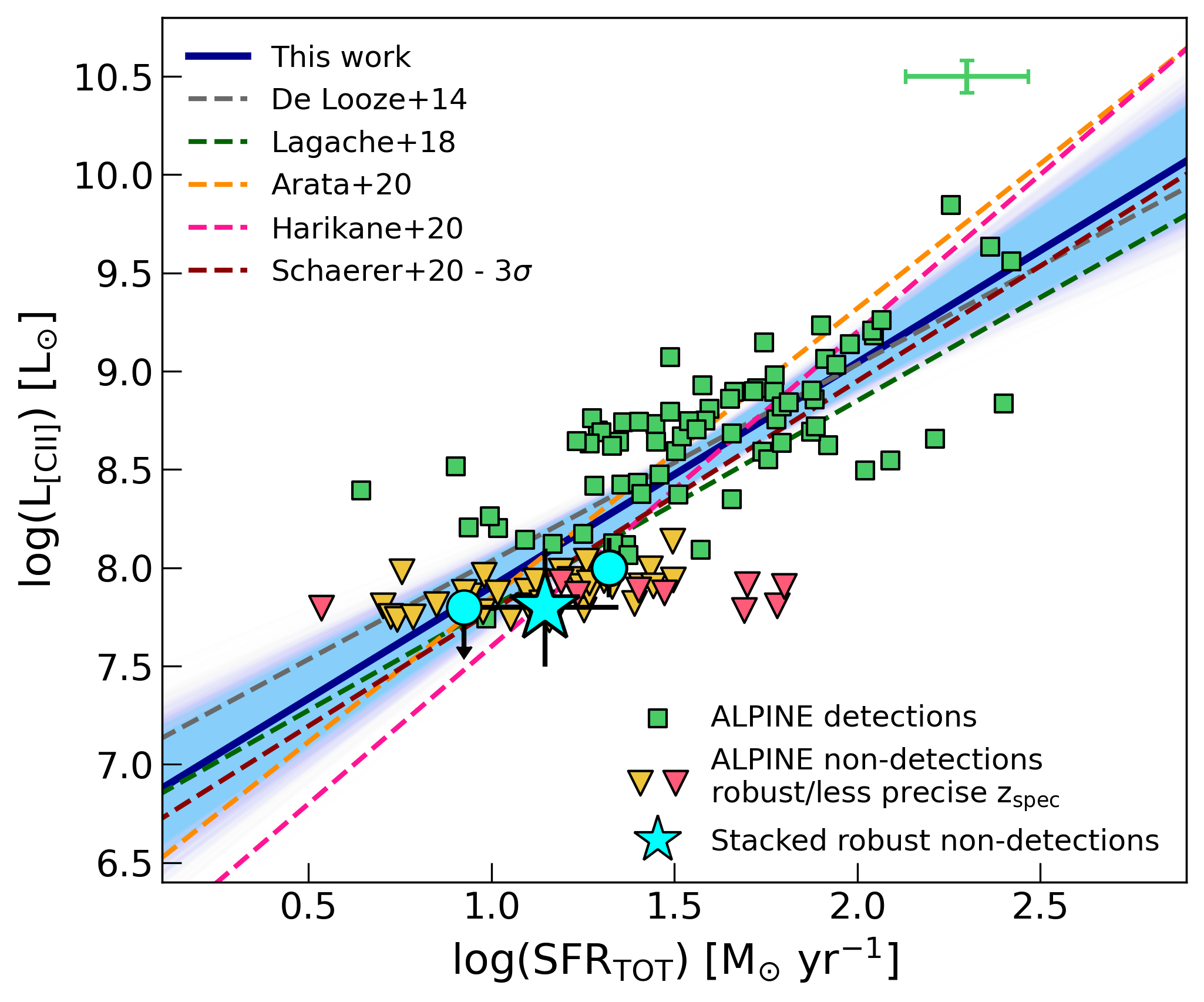}
	\end{center}
    \caption{[CII] luminosity as a function of total SFR (UV+FIR; see text) for the ALPINE detections (green squares) and non-detections (as $3\sigma$ upper limits; yellow triangles). The red markers represent the [CII]-undetected galaxies with less precise $z_\mathrm{spec}$ (see Section \ref{sec:data}). The green errobars in the upper right corner show the average uncertainties of [CII] detections on both L$_\mathrm{[CII]}$ and SFR. The cyan big star shows the result of this work for the stacked non-detections, with the error on L$_\mathrm{[CII]}$ computed as described in Section \ref{sec:stacking}. The cyan circles are the [CII] upper limit and detection found in the low- and high-SFR bins introduced in the text, respectively. We also report different L$_\mathrm{[CII]}$-SFR relations from the literature: the local relation by \cite{DeLooze14} (dashed gray line), the predicted relation by the models of \cite{Lagache18} at $z\sim5$ (dashed green line), the predicted relation by \cite{Arata20} for galaxies at $z>6$ (dashed yellow line), the fitted relation by \cite{Harikane20} on $z>6$ galaxies (dashed pink line), the latest results obtained by \cite{Schaerer20} at $z\sim5$ with ALPINE considering non-detections as $3\sigma$ upper limits (dashed red line). Finally, the solid blue line and shaded area report our best-fit and uncertainties to the [CII] detections + stacked robust non-detections.}
    \label{fig:[CII]sfr}
\end{figure}

By taking advantage of the [CII] luminosity computed from the stack of the ALPINE non-detections, we explore the relation between $\mathrm{L_{[CII]}}$ and SFR in these galaxies in order to compare them to the results obtained for the combined [CII] detections and non-detections upper limits \citep{Schaerer20}. 

Following \cite{Schaerer20}, we report in Figure \ref{fig:[CII]sfr} the [CII] luminosities and SFRs of the ALPINE detections and non-detections (as 3$\sigma$ upper limits; see also Figure 4 by \citealt{Schaerer20}). Contrarily to Figure \ref{fig:MS} in which we show the SFRs obtained through SED-fitting (for consistency with the stellar mass values obtained with the same procedure; \citealt{Faisst20}), we use in this case the total SFRs obtained as $\mathrm{SFR_{TOT} = SFR(UV)+SFR(IR)}$, where SFR(UV) comes from the observed UV absolute magnitude at 1500~$\AA$, and SFR(IR) is obtained both through the ALMA continuum measurements and, for galaxies undetected in continuum, through predictions of the IRX-$\beta$ relation by \cite{Fudamoto20}\footnote{It is worth to specify that none of our 34 [CII] non-detections is detected in continuum. Therefore, the SFR(IR) for all of these galaxies comes from the IRX-$\beta$ relation by \cite{Fudamoto20}.}, obtained from a stacking analysis of all ALMA continuum images, including both individual detections and non-detections. In this way, we are able to compare our results with those previously found in ALPINE by \cite{Schaerer20} and with other $\mathrm{L_{[CII]}}$-SFR relations already present in the literature. Among these, we show: i) the local $\mathrm{L_{[CII]}}$-SFR relation found by \cite{DeLooze14} for a sample of low-z HII/starburst galaxies\footnote{As this relation is based on a \cite{Kroupa03} IMF, we scaled it to a \cite{Chabrier03} IMF by dividing the SFR by a factor 1.06 (e.g., \citealt{Madau14}), for consistency with other measurements from the literature.}; ii) the predicted relation found by \cite{Lagache18} at $z=5$; iii) the predictions from simulations for $z>6$ galaxies by \cite{Arata20}; iv) the relation found by \cite{Harikane20} for galaxies observed in [CII] at $6<z<9$; iv) the relation fitted to the ALPINE data only (including 3$\sigma$ upper limits on non-detections) by \cite{Schaerer20}. Furthermore, we show our results from the stacking of ALPINE [CII] non-detections (i.e., $\mathrm{log(L_\mathrm{[CII]}}/\mathrm{L_{\odot}}) = 7.8\pm0.3$), adopting their mean $\mathrm{SFR_{TOT}}$ (i.e., $\mathrm{log(SFR_{TOT,mean}/M_{\odot}~yr^{-1}})=1.14\pm0.20$\footnote{The error on $\mathrm{SFR_{TOT,mean}}$ is computed from the 16th and 84th percentiles of the $\mathrm{SFR_{TOT}}$ distribution of the [CII] non-detections.}, that is also in good agreement with the upper limit found by \cite{Bethermin20} by stacking the ALMA continuum of the [CII] non-detections. For further information, see their Section 7.5). For comparison, we also report the [CII] upper-limit and detection found in the low- and high-SFR bin respectively, as described in Section \ref{sec:stacking}. 

Our $\mathrm{L_\mathrm{[CII]}}$ value is consistent with the $3\sigma$ upper limits found by \cite{Bethermin20} and used by \cite{Schaerer20} in their analysis. This result suggests that the [CII] non-detections are not drawn from a different population of galaxies with respect to the ALPINE detections. Rather, these are galaxies lying on the bottom-left region of the $z\sim5$ main-sequence with lower SFRs and stellar masses, and possibly with fainter [CII] emission (see also Section \ref{sec:discussion} for possible caveats causing low [CII] content in high-$z$ galaxies).

We fit the combined [CII] luminosity and SFR from the stacking of non-detections with the ALPINE [CII]-detected galaxies with a linear relation of the form
\begin{equation}
    \mathrm{log(L_\mathrm{[CII]}/L_\odot)} = a + b \times \mathrm{log(SFR/M_{\odot}~yr^{-1}}),
    \label{eq:fit}
\end{equation}
where $a$ and $b$ are the intercept and slope of the relation, respectively. For consistency with \cite{Schaerer20}, we use the \texttt{linmix} package\footnote{{\url{https://github.com/jmeyers314/linmix}}.} by \cite{Kelly07} which makes use of a Bayesian approach to account for measurement errors in both variables in linear regressions. The errors on the [CII] luminosities are taken from \cite{Bethermin20} for the ALPINE detections, and from the stacking for the non-detected sources. Regarding the individual uncertainties on the SFRs, we propagate the errors of SFR(UV) and SFR(IR) on the total SFR (obtaining on average $\sim0.2$~dex for both detections and non-detections).

With our stacked non-detections, we obtain a best-fit relation with $a=6.76\pm0.17$ and $b=1.14\pm0.11$. This is consistent with the slope obtained from local galaxies ($b=1.00\pm0.04$; \citealt{DeLooze14}), and in agreement with that previously found by \cite{Schaerer20} considering detections and $3\sigma$ upper limits on non-detections (i.e., $b=1.17 \pm 0.12$). Table \ref{tab:relations} summarizes the parameters describing the L$_\mathrm{[CII]}$-SFR relations found in the literature and in this work.

\begin{table*}
\caption{Summary of L$_\mathrm{[CII]}$-SFR relations from the literature and from this work, as parameterized in Equation \ref{eq:fit}.}
\label{tab:relations}
\begin{center}
\begin{tabular}{c c c c c}
\hline
\hline
Literature & Sample & Redshift & \textit{a} & \textit{b}\\
(1) & (2) & (3) & (4) & (5)\\
\hline
\cite{DeLooze14} & HII/starburst & $<0.5$ & $7.06\pm0.33$ & $1.00\pm0.04$\\
\cite{Lagache18} & \texttt{G.A.S. + CLOUDY} & 4 - 6 & $6.75\pm0.07$ & $1.05\pm0.07$\\
\cite{Arata20} & \texttt{GADGET-3 + ART$^{2}$} & >6 & 6.38 & 1.47\\
\cite{Harikane20} & LBGs/SMGs & 6 - 9 & 6.00 & 1.60\\
\cite{Schaerer20} & ALPINE ($3\sigma$ limits) & 4 - 6 & $6.61\pm0.20$ & $1.17\pm0.12$\\
This work & ALPINE det + stacked non-det & 4 - 6 & $6.76\pm0.17$ & $1.14\pm0.11$\\ 
\hline
\end{tabular}
\end{center}
\end{table*}

\section{Discussion}\label{sec:discussion}
A large scatter in the $\mathrm{L_{[CII]}}$-SFR relation is in place for sources at $z>4$ (i.e., $\gtrsim0.4$~dex, that is $\sim2$ times larger than the intrinsic dispersion of local galaxies; \citealt{Carniani18a,Schaerer20}). Such a scatter is produced by the multitude of [CII] detections and non-detections now available in the high-$z$ Universe, and could be due both to different physical conditions in the ISM of distant galaxies, or to systematic in the [CII] and SFR derivation.

\cite{Jolly21} analyzed a sample of 52 gravitationally lensed galaxies at $z\gtrsim6$ as part of the ALMA Lensing Cluster Survey (ALCS) searching for [CII] emission through spectral stacking. They found no [CII] detection in their sample, providing $3\sigma$ upper limits on the [CII] luminosity of $\mathrm{log(L_\mathrm{[CII]}}/\mathrm{L_{\odot}}) < 7-8$, depending on the considered sub-sample (e.g., sources with high SFR, or with secure spectroscopic redshift) and stacking method (i.e., weighted mean and median stacking)\footnote{It is worth saying that \cite{Jolly21} did not find a significant difference between their mean and median stacked cubes. Rather, most of the discrepancy in their two methods resides in the difference between the mean and median magnification of their sample.}. Their findings are comparable\footnote{Although the galaxies in \cite{Jolly21} cover wider SFR and stellar mass ranges (reaching $\mathrm{log(SFR/M_{\odot}~yr^{-1}})<-0.5$), a fair comparison with our sample of [CII] non-detections is still feasible if considering their high-SFR sub-sample which nicely matches our SFR distribution, and for which they found $\mathrm{log(L_\mathrm{[CII]}}/\mathrm{L_{\odot}}) < 7.7-7.9$, depending on the stacking method used.} with (or lower than) our average [CII] luminosity from the ALPINE non-detections (i.e., $\mathrm{log(L_\mathrm{[CII]}}/\mathrm{L_{\odot}}) = 7.8\pm0.3$). However, these authors claim that the [CII] upper limits they found could be underestimated because some potential biases in the analyzed galaxy sample. First, they do not account for the spectroscopic offset between Ly$\alpha$ and [CII] that, as mentioned in Section \ref{sec:stacking}, could alter the stacked line profile and lower the average amount of [CII] luminosity arising from non-detections. Astrometric offsets between ALMA and ancillary data could affect the [CII] luminosity estimate, as well. This issue was well-studied by many authors in the literature (e.g., \citealt{Dunlop17,Franco18}), and in ALPINE by \cite{Faisst20} and \cite{Fujimoto20}, who found a typical shift between [CII] and UV rest-frame emission of 0.15'' (corresponding to $\sim1$~kpc at $z=5$). We overcome this problem by considering a fixed aperture of 1'' of radius to extract the ALMA spectra of the [CII] non-detections (which is also $\sim3$ times larger than the typical [CII] size of individual detections in ALPINE, i.e., 2.1~kpc). The surface brightness distribution of the [CII] emission and its FWHM could also reduce the [CII] luminosity and cause non-detections. For instance, \cite{Fujimoto20} found that [CII] emission is typically 2-3 times more extended than rest-frame UV emission, which could result in a large underestimation of the total [CII] flux, depending on the resolution of the observations (e.g., \citealt{Carniani20}). \cite{Kohandel19} studied instead the effect of FWHM and inclination of the galaxy on the [CII] detectability, finding that an edge-on galaxy with a large line width (i.e., FWHM $>400$~km~s$^{-1}$) could result in a non-detection, biasing the [CII] luminosity estimates in high-$z$ sources. Finally, \cite{Jolly21} noticed that their results may be lowered by the presence of a large number of Ly$\alpha$ emitters in their sample, that could in principle have small content of dust, resulting in low [CII] and FIR emission. This issue has been minimized in the ALPINE sample by a heterogeneous pre-selection of the spectroscopically observed
galaxies done with a variety of diverse selection methods \citep{Faisst20}. In addition, \cite{Cassata20} found that 44\% of the ALPINE sources are Ly$\alpha$ emitters\footnote{We follow here \cite{Cassata20}, for which a Ly$\alpha$ emitter is defined as a source with rest-frame equivalent width $\mathrm{EW_0(Ly\alpha)>25~\AA}$.}, with the probability of detecting [CII] not strongly dependent on the presence of bright Ly$\alpha$ emission.

Apart from the above caveats, the little [CII] content found by \cite{Jolly21} in their galaxies could also be due to low metallicity levels. Indeed, many simulations show that low metallicity in high-$z$ galaxies could result in a lower Carbon abundance, and stronger radiation field that could decrease the [CII] content (e.g., \citealt{Vallini15,Ferrara19,Vallini20}). About this, it is interesting to note that our L$_\mathrm{[CII]}$-SFR slope (i.e., $b=1.14 \pm 0.11$) is also consistent with that found by \cite{DeLooze14} for metal-poor dwarf galaxies in the local Universe (i.e., $b\sim1.25$). This is in line with the results from \cite{Faisst20} who compared the H$\alpha$ (which is a good tracer of the star-formation properties of galaxies) and [CII] luminosities for a sub-sample of the ALPINE galaxies. They found that bright [CII] galaxies are in good agreement with the local relation between H$\alpha$ and [CII] found by \cite{DeLooze14}. However, for lower [CII] luminosities ($<5\times10^{8}~\mathrm{L_{\odot}}$, as those probed by the [CII] non-detections), the galaxies seem to be more consistent with the relation found for local metal-poor dwarf galaxies, although with a large scatter. The possible effect of metallicity on [CII] is also suggested by the results of \cite{Faisst20} who noticed that the strength of [CII] emission increases along the main-sequence, with a significant decrease of [CII] content in galaxies with $\mathrm{log(M_{*}/M_{\odot})<9.3}$ and $\mathrm{log(SFR/M_{\odot}yr^{-1})<1}$ (i.e., in [CII]-undetected sources). This further suggests that the metallicity of galaxies (along with the strength of their [CII] emission) could play an important role in the derivation of the L$_\mathrm{[CII]}$-SFR relation \citep{Vallini15,Olsen17,Lagache18,Narayanan18,Ferrara19}. Future investigations of this topic will be possible thanks to the forthcoming near- and mid-infrared observations of the James Webb Space Telescope, through which we will be able to provide measurements of the metallicity content of distant galaxies.

To better understand the relation between [CII] and SFR in normal $z\sim5$ galaxies, we compare our result with those of different simulations. \cite{Lagache18} used the semi-analytical model \texttt{G.A.S.} (Galaxy Assembly from dark-matter Simulations) in combination with the photoionization code \texttt{CLOUDY} \citep{Ferland13,Ferland17} to predict the [CII] luminosity of a large number of galaxies at $z\sim5$. They found an average relation with a slope in agreement with the one by \cite{DeLooze14} for HII/starburst galaxies at low redshift, but that is dependent on several parameters, such as the metallicity of the galaxies and the intensity of their interstellar radiation field. More recently, \cite{Arata20} combined cosmological hydrodynamic simulations performed with the \texttt{GADGET-3} code \citep{Springel05}, with the All-wavelength Radiative Transfer with Adaptive Refinement Tree \texttt{(ART$^{2}$)} code \citep{Li08,Yajima12} to predict the relation between [CII] and SFR for galaxies well within the Reionization epoch, at $z>6$. They found a steep slope (i.e., $b=1.47$) of the relation, suggesting that the deviation from the local Universe is caused by changes in the distribution of neutral gas in high-$z$ galaxies. Observationally, similar results were obtained by \cite{Harikane20} who found a very steep slope (i.e., $b=1.6$) by analyzing a sample of $6<z<9$ LBGs and SMGs. They used \texttt{CLOUDY} to investigate the physics below their results, finding that such a slope (that is mainly caused by galaxies with low $\mathrm{L_{[CII]}/SFR}$ ratios at low SFRs) could be produced by a high ionization parameter (as a result of a strong correlation between gas-phase metallicity and SFR, or of a bursty star formation in galaxies) and/or by a low PDR covering fraction (that is where most of the [CII] emission comes from).

Our results differ significantly from those obtained by \cite{Harikane20} and simulated by \cite{Arata20}. As suggested by \cite{Schaerer20}, this could be due to the fact that we use uniform estimation of the total SFR based on rest-frame UV and FIR measurements, instead of using SED-based SFRs which are typically affected by the choice of the star-formation histories, dust-attenuation curves, and stellar populations (e.g., \citealt{Carniani20}). On the other hand, \cite{Harikane20} claimed that the difference in the method of computing SFRs does not significantly affect the estimate of the L$_\mathrm{[CII]}$-SFR relation, rather the different slope could be ascribed to a change in the properties of galaxies from $z\sim5$ to higher redshift, or to a diverse selection of the parent sample. However, in order to put stronger constraints on the [CII] properties in the early Universe, further observations at $\mathrm{log(SFR/M_{\odot}yr^{-1})<1}$ are needed. 

Finally, it is worth noting that, both in the stack and in the fit, we do not include 9 non-detections having less precise spectroscopic redshifts than those from the other galaxies in the sample (see Section \ref{sec:UV_data}). As evidenced in Figure \ref{fig:[CII]sfr}, some of these sources show the largest deviation from the derived L$_\mathrm{[CII]}$-SFR relation, suggesting the possible presence of the so-called \lq[CII]-deficit\rq{} (e.g., \citealt{Malhotra01,Vallini15,Lagache18,Harikane20}). However, these galaxies are likely affected by a poor estimate of their spectroscopic redshifts with respect to the other ALPINE non-detections. Indeed, for the analysis undertaken in this work, an inaccurate derivation of $z_\mathrm{spec}$ could induce an not physical offset between the rest-frame UV spectroscopic redshift and the systemic one as traced by the [CII] line that we are not able to correct based on the observed $\Delta v_\mathrm{Ly\alpha}$ distribution. At worst, the expected emission line could be moved outside of the ALMA spectral window of observation. In this scenario, we claim that there is no evidence of [CII]-deficit within the ALPINE sample, as also suggested by \cite{Schaerer20}. 



\section{Summary and conclusions}\label{sec:summary}
It is now well established that, in the local Universe, a linear relation between the [CII] luminosity and the SFR of galaxies is in place (e.g., \citealt{DeLooze14}). Whether this relation holds at earlier epochs is still debated. At high redshifts, the number of [CII] detections is increasing, but there are still few constraints from the low SFR, stellar mass regime, which are fundamental to properly characterize the $\mathrm{L_{[CII]}}$-SFR relation over cosmic time. Indeed, only a handful of low-SFR, high-$z$ sources have been detected (or undetected) so far in [CII], most of them as strongly lensed galaxies (e.g., \citealt{Knudsen16,Carniani17,Laporte19,Fujimoto21,Jolly21,Laporte21}).

In this work, we perform a spectral stacking of 34 [CII]-undetected galaxies at $z\sim5$, as part of the ALPINE survey, taking into account the typical observed rest-frame UV-FIR spectral offset between the Ly$\alpha$ and [CII] lines \citep{Cassata20}. The stack reveals a [CII] detection at $\sim5.1\sigma$, providing one of the few constraints on the L$_\mathrm{[CII]}$-SFR relation at high redshift and in the low-SFR regime, and resulting in a line luminosity of $\mathrm{log(L_\mathrm{[CII]}}/\mathrm{L_{\odot}}) \sim 7.8$. By fitting the average [CII] luminosity from the stacking of the non-detections with that from the individual ALPINE [CII]-detected galaxies as a function of their total SFRs, we find a linear relation that is comparable with the local one ($1.14\pm0.11$ from this work against $1.00\pm0.04$ from \citealt{DeLooze14}) and in agreement with the previous results by \cite{Schaerer20}, suggesting that [CII] is still a good tracer of star formation in the early Universe. In this respect, we do not even find evidence of the so-called [CII]-deficit, that could steepen the L$_\mathrm{[CII]}$-SFR slope as found in previous works (e.g., \citealt{Harikane20}).

However, further and deeper observations are needed in order to confirm these results, especially in the low SFR regime, where the L$_\mathrm{[CII]}$-SFR relation is consistent from the spatially-resolved and the entire galaxy scales, and where only a handful of strongly lensed galaxies have been detected so far at $\mathrm{log(SFR/M_{\odot}~yr^{-1}}) \lesssim 0.5$ (e.g., \citealt{Knudsen16,Fujimoto21}). Current simulations provide different results at high redshift, suggesting that many physical mechanisms could be in place in the ISM of distant galaxies (e.g., \citealt{Lagache18,Arata20}). Further constraints, as the one provided in this work, on the low L$_\mathrm{[CII]}$ and SFR tail of normal high-$z$ SFGs could serve as input for such cosmological simulations in order to shed light on the physics of [CII] in the early Universe.


\begin{acknowledgements}
We warmly thank the anonymous referee for her/his careful reading of our paper and for the constructive comments and useful suggestions that have contributed to improve the quality of the work. This paper is based on data obtained with the ALMA Observatory, under Large Program 2017.1.00428.L. ALMA is a partnership of ESO (representing its member states), NSF(USA) and NINS (Japan), together with NRC (Canada), MOST and ASIAA (Taiwan), and KASI (Republic of Korea), in cooperation with the Republic of Chile. The Joint ALMA Observatory is operated by ESO, AUI/NRAO and NAOJ. M.R. acknowledges support from the Narodowe Centrum Nauki (UMO-2020/38/E/ST9/00077). M.T. acknowledges the support from grant PRIN MIUR 2017 20173ML3WW 001. This work was supported by the Programme National Cosmology et Galaxies (PNCG) of CNRS/INSU with INP and IN2P3, co-funded by CEA and CNES. The Cosmic Dawn Center is funded by the Danish National Research Foundation under grant No. 140. This project has received funding from the European Union’s Horizon 2020 research and innovation program under the Marie Sklodowska-Curie grant agreement No. 847523 ‘INTERACTIONS'. G.C.J. acknowledges ERC Advanced Grant 695671 ``QUENCH'' and support by the Science and Technology Facilities Council (STFC). M.B. gratefully acknowledges support by the ANID BASAL project FB210003 and from the FONDECYT regular grant 1211000. This paper is dedicated to the memory of Olivier Le Fèvre, PI of the ALPINE survey.
\end{acknowledgements}

%
%

\bibliographystyle{aa} 
\bibliography{aanda.bib} 

\end{document}